\documentstyle[prb,aps,multicol]{revtex}

\begin{document}

\preprint{Preprint}
\draft{}

\title{Avalanche dynamics, surface roughening and self-organized
criticality - experiments on a 3 dimensional pile of rice.}

\author{C.~M.~Aegerter, R.~G\"unther, and R.~J.~Wijngaarden}

\address{Division of Physics and Astronomy, Faculty of Sciences,
Vrije Universiteit, De Boelelaan 1081, 1081HV Amsterdam, The Netherlands}

\date{\today}
\maketitle
\widetext

\begin{abstract}
We present a two-dimensional system which exhibits features of
self-organized criticality. The avalanches which occur on the
surface of a pile of rice are found to exhibit finite size scaling
in their probability distribution. The critical exponents are
$\tau$ = 1.21(2) for the avalanche size distribution and $D$ =
1.99(2) for the cut-off size. Furthermore the geometry of the
avalanches is studied leading to a fractal dimension of the active
sites of $d_B$ = 1.58(2). Using a set of scaling relations, we can
calculate the roughness exponent $\alpha = D - d_B$ = 0.41(3) and
the dynamic exponent $z = D(2 - \tau)$ = 1.56(8). This result is
compared with that obtained from a power spectrum analysis of the
surface roughness, which yields $\alpha$ = 0.42(3) and $z$ =
1.5(1) in excellent agreement with those obtained from the scaling
relations.
\end{abstract}
\pacs{DRAFT VERSION: NOT FOR DISTRIBUTION}

\begin{multicols}{2}
\section{Introduction}

The concept of self-organized criticality (SOC) \cite{SOC}
presents a simple way of modelling slowly-driven out of
equilibrium systems. The interesting natural systems thought to
exhibit SOC, such as rain-fall \cite{ole}, earthquakes
\cite{earthquake}, economic markets \cite{economy}, biological
evolution \cite{bio}, or the brain \cite{brain}, are difficult to
study in a controlled experiment, and more simple systems have to
be found with which the predictions of SOC can be tested
quantitatively. Among the first toy systems studied, were
sand-piles, however it has been shown that due to the appearance
of e.g. kinetic effects, real sand does not behave as an SOC
system \cite{jaeger}. Another system, that has shown indications
of SOC in a controlled environment is the dynamics of vortices in
type-II superconductors. Here, power-law scaling has been observed
in the past \cite{field}, however not all authors have found
avalanche scaling \cite{aval}. This is probably due to the fact
that internal avalanches are not measured in the usual setup
\cite{feder}. Recently Behnia {\em et al.} \cite{behnia} have
measured internal avalanches using arrays of Hall-probes, where
they do indeed find power-law behavior of the avalanches. The
ultimate hall-mark of SOC, however, is the observation of finite
size-scaling, indicating a true critical dynamics. So far, there
are only two experimental systems, a one dimensional (1d) pile of
rice \cite{Frette} or a 1d pile of steel balls with a random
arrangement at its bottom layer \cite{ernesto2}, which have been
shown to exhibit finite size scaling in their avalanche
distributions.

The critical exponents obtained from a finite size scaling
analysis, would yield more information on a SOC system, as was
discussed by Paczuski {\em et al.} \cite{pacz} for models in a
stationary critical state. A system evolving through avalanches,
which are distributed according to a power-law, will also show
roughening dynamics of its surface. This can be seen as an
indication to the origin of the abundance of fractal or
self-affine structures in nature. Roughening dynamics has been
studied extensively in the past, including the study of interfaces
in porous media \cite{porous}, the growth of bacterial colonies
\cite{bact}, the slow combustion \cite{jussi} or the rupture
\cite{paper} of paper . These systems are well characterized
experimentally and in many cases, the interface dynamics can be
modelled analytically by the Kardar-Parisi-Zhang (KPZ) equation
\cite{KPZ}. In one spatial dimension, the KPZ equation is solved
exactly for the case of white noise, such that good comparison
between experiment and theory is possible. However, e.g. in the
burning of paper, there are avalanches observed in the propagation
of the front, which are reminiscent of SOC dynamics \cite{jussi2}.
This has recently been addressed by Alava {\em et al.}
\cite{alava}, who proposed a mapping between SOC models and the
KPZ equation similar to that of Paczuski and Boettcher
\cite{boettcher}. With this mapping, they have been able to obtain
SOC from the KPZ equation, with the proper kind of noise
\cite{alava}.

Moreover, roughening processes have also been found in the
experimental SOC systems discussed above. The profile of a 1d rice
pile is a rough interface \cite{rice}, as is the front of magnetic
flux penetrating a thin film superconducting sample \cite{radu}.
However, the universal scaling relations derived by Paczuski {\em
et al.} \cite{pacz} have not yet been tested experimentally. But
the fact that both roughening phenomena and avalanche dynamics can
be observed in the same experimental system, such as a 3d
rice-pile, means that the universal connections between SOC and
surface roughening can be tested experimentally.

Here we present a two-dimensional (2d) system, the surface of a 3d
pile of rice, showing both roughening behavior and avalanche
dynamics. The avalanche dynamics is studied in terms of the
avalanche size distribution for different sizes of the field of
view, $L$. This allows the observation of finite size scaling and
the determination of the critical exponents $D$, describing the
dependence of the cutoff scale on $L$, and $\tau$, the exponent of
the avalanche size distribution. The roughening behavior is
studied via the power spectrum of the surface \cite{power}, in
both space and time, resulting in a determination of the roughness
and growth exponents respectively \cite{barabasi}. The connection
between these two phenomena is shown from the derivation of
scaling relations between the different exponents, where we obtain
excellent agreement using the experimentally determined values.

The experimental setup for the growth of the rice-pile, as well as
the reconstruction method used to determine the structure of the
pile and the size of the avalanches is discussed in section II.
The avalanche size distributions, and their finite size scaling is
presented in section IIIA, together with the determination of the
critical exponents. In section IIIB, the surface roughness is
analyzed and the necessary techniques are briefly introduced.
Still in section IIIB, the scaling relations between the roughness
exponents and the critical exponents are introduced. These results
are also put into a wider context and compared with results from
KPZ roughening systems \cite{2dKPZ}.

\section{Experimental Details}

\begin{figure}
\input{epsf}
\epsfxsize 8cm 
\centerline{\epsfbox{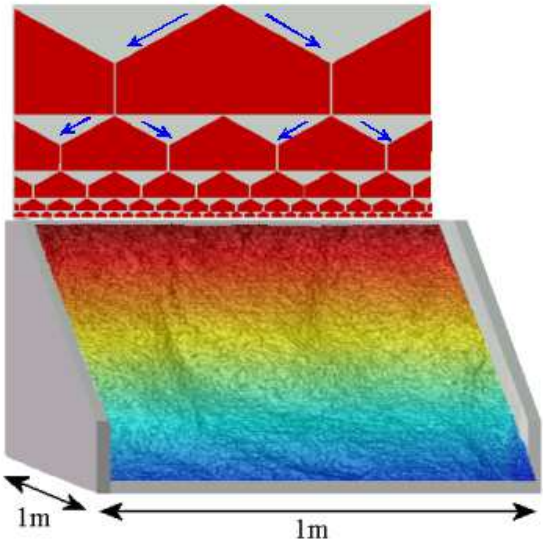}} \caption[~]{A schematic image of
the setup. The distribution board can be seen on top, where rice
is dropped from a single point and subsequently divided into even
compartments. Within the wooden box bounding the rice pile, a
reconstruction of its surface is shown, as it is used in the
further analysis.} \label{dist}
\end{figure}

The experiments were carried out on long grained rice with
dimensions of typically $\sim$2x2x7 mm$^3$, similar to rice A of
ref.~\cite{Frette}. The pile was grown from a uniform line source
1 m wide. This uniform distribution was achieved via a custom
built mechanical distributor based on a nail-board producing a
binary distribution \cite{rad}. The actual setup consists of a
board with an arrangement of triangles as shown in
Fig.~\ref{dist}. This means that the possible pathways of the rice
are continuously split at each level, such that at the bottom we
end up with a row of 64 uniformly distributed compartments. The
uniformity of the distribution was measured to be of the order of
5$\%$.

At the bottom of the distributor, the grains are slowed down by a
sheet of plastic before they hit the top of the pile. The rice is
fed to the top of the distribution board, from a point source,
which drops rice at an average rate of $\sim$5g/s. This
corresponds to 1500 grains per image, which means that over the
length of the line there are 1 to 2 grains dropped per place. This
rate is uniform over the time scale between two images.

\begin{figure}
\input{epsf}
\epsfxsize 8cm 
\centerline{\epsfbox{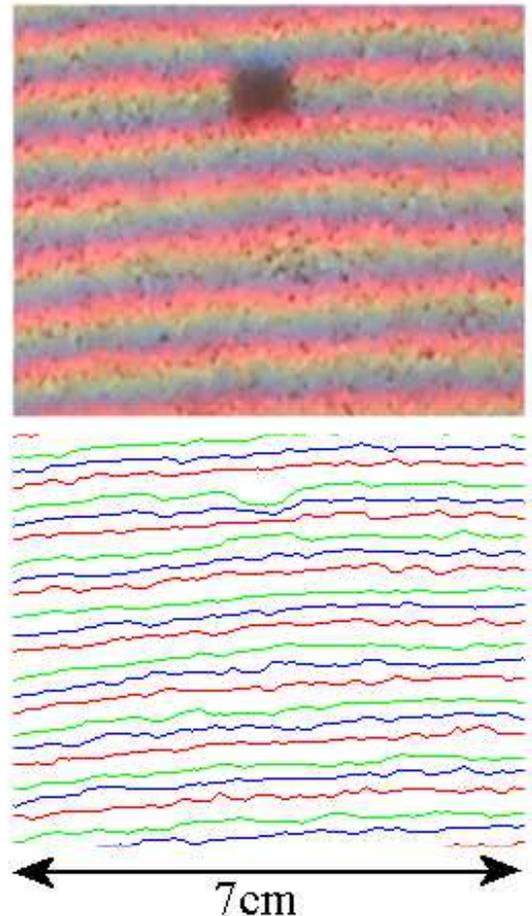}} \caption[~]{A raw image of the
rice pile with the identified lines. From the distortions of the
lines, the surface geometry is obtained. The dark spot in the
image indicates a reference point.} \label{lines}
\end{figure}

Once a rice pile is grown, we measure the further evolution of the
surface coordinates using a specially developed real-time
technique based on the projection of a set of lines. In order to
increase the spatial resolution in the direction of growth, we use
a set of lines in the base colors (red, green and blue), which can
be easily separated digitally for the analysis. The lines are
projected at a right angle to the average surface of the pile and
observed at 45$^\circ$ to both the direction of projection as well
as the surface. This is illustrated in Fig.~\ref{lines}, where a
raw image is shown together with the lines extracted from it. From
the distortions of the lines observed in this way, the coordinates
in 3d space of the pile surface can be calculated by simple
geometry \cite{rad}. With this reconstruction technique, the 3d
coordinates of the whole field of view can be determined with an
accuracy of 2-3 mm, as we have checked on a number of test
surfaces. Using both structured and smooth surfaces, the
resolution and accuracy were determined independently.  Since the
resolution is roughly matched with the size of the grains, the
method is well suited for the present purpose.

In a single experimental run, the growth is studied for a period
of $\sim$4 hours, with a picture taken every 30 s. Thus an
experiment consists of $\sim$ 480 images. The pictures are taken
with a high resolution digital camera, having a resolution of
2048x1536 pixels. For every time step, the surface structure is
reconstructed, which gives information about the roughening
process, as well as the avalanche dynamics.

\begin{figure}
\input{epsf}
\epsfxsize 8cm 
\centerline{\epsfbox{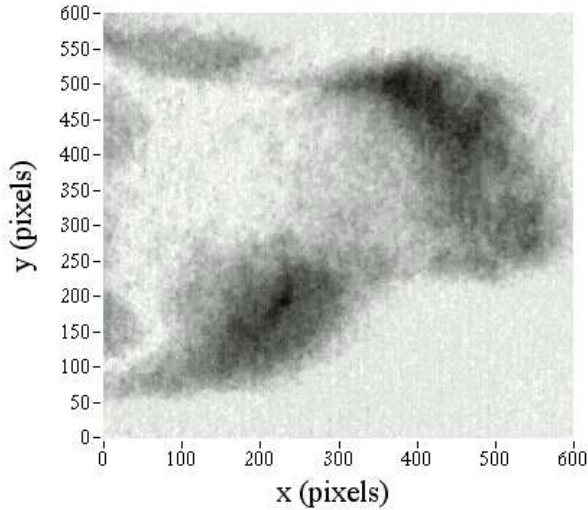}} \caption[~]{A snapshot of an
avalanche, given by the height difference of two subsequent
images. In the picture, the height is given by the grey-scale with
white corresponding to $\Delta h$ = 0 and black to $\Delta h$ = 15
mm. Integrating the height difference $\Delta h(x,y)$ over the
area gives the size of the avalanche in terms of a volume, or in
terms of the number of moved grains from the volume occupied by a
grain $V_{grain}$ = 35 mm$^3$.} \label{diff}
\end{figure}

The size and shape of the avalanches can be determined from the
height difference of the surface between two consecutive images.
The overall growth of the pile is subtracted from this difference,
however this correction is negligible and does not influence the
results. This is shown in Fig.\ref{diff}, where the height
difference $\Delta h(x,y)$ is shown for a medium size avalanche.
This allows the study of internal avalanches instead of just the
off-edge ones, which has proven to be important in previous
studies (for instance on superconducting vortex avalanches
\cite{behnia} and 1d piles of rice \cite{Frette}). In order to
obtain the size of such an avalanche, $|\Delta h|$ is integrated
over the area, which yields the displaced volume, $\Delta V$,
corresponding to the size of the avalanche in mm$^3$. In order to
use natural units, we will in the following measure the size of
the avalanche in terms of the number of toppled grains, which is
$s = \Delta V/V_{grain}$, where $V_{grain} \simeq$ 35 mm$^3$ is
the volume of a single grain of rice.

The data discussed in this article was obtained in three separate
experiments, with a total of 1330 images.

\section{Results}

In this section the experimental results are presented. We first
show that there is finite size scaling in the avalanche
statistics, which indicates the appearance of a critical state in
the system. Second, we characterize the surface roughness of the
rice pile in 2d, both in space and time. The two characteristic
exponents, $\alpha$ for the spatial behavior, and $z$ for the
dynamics \cite{barabasi} can also be obtained from scaling
relations of the critical exponents of the avalanche statistics
and are in excellent agreement with those determined from a
standard power-spectrum analysis.

\subsection{Avalanche Dynamics}

\begin{figure}
\input{epsf}
\epsfxsize 9cm 
\centerline{\epsfbox{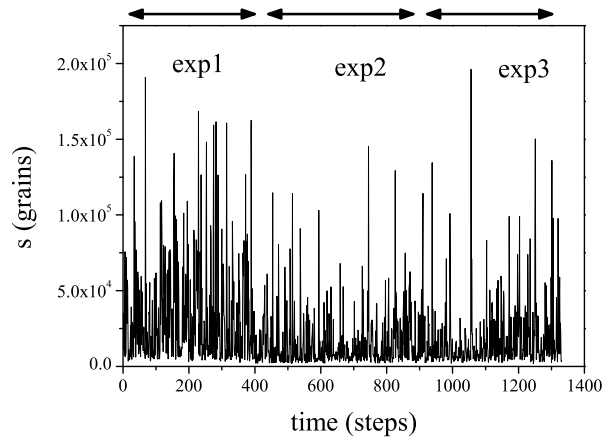}} \caption[~]{The time evolution of
the integrated height difference for the three experimental runs
studied. As can be seen, times of relative rest are punctuated by
large avalanches, which appear intermittently. When studying the
distribution of avalanche sizes (see Fig.~\ref{hist}), it is found
that there is no intrinsic size to the avalanches, but that they
are distributed according to a power-law. The different
experimental runs are indicated by arrows.} \label{time}
\end{figure}

The time dependence of the displaced volume, $\Delta V$, in each
time step is shown in Fig.~\ref{time}, for all three experiments
discussed here. In the figure, the data are shown for avalanche
sizes integrated over the total field of view of 600x600 mm$^2$.

\begin{figure}
\input{epsf}
\epsfxsize 9cm 
\centerline{\epsfbox{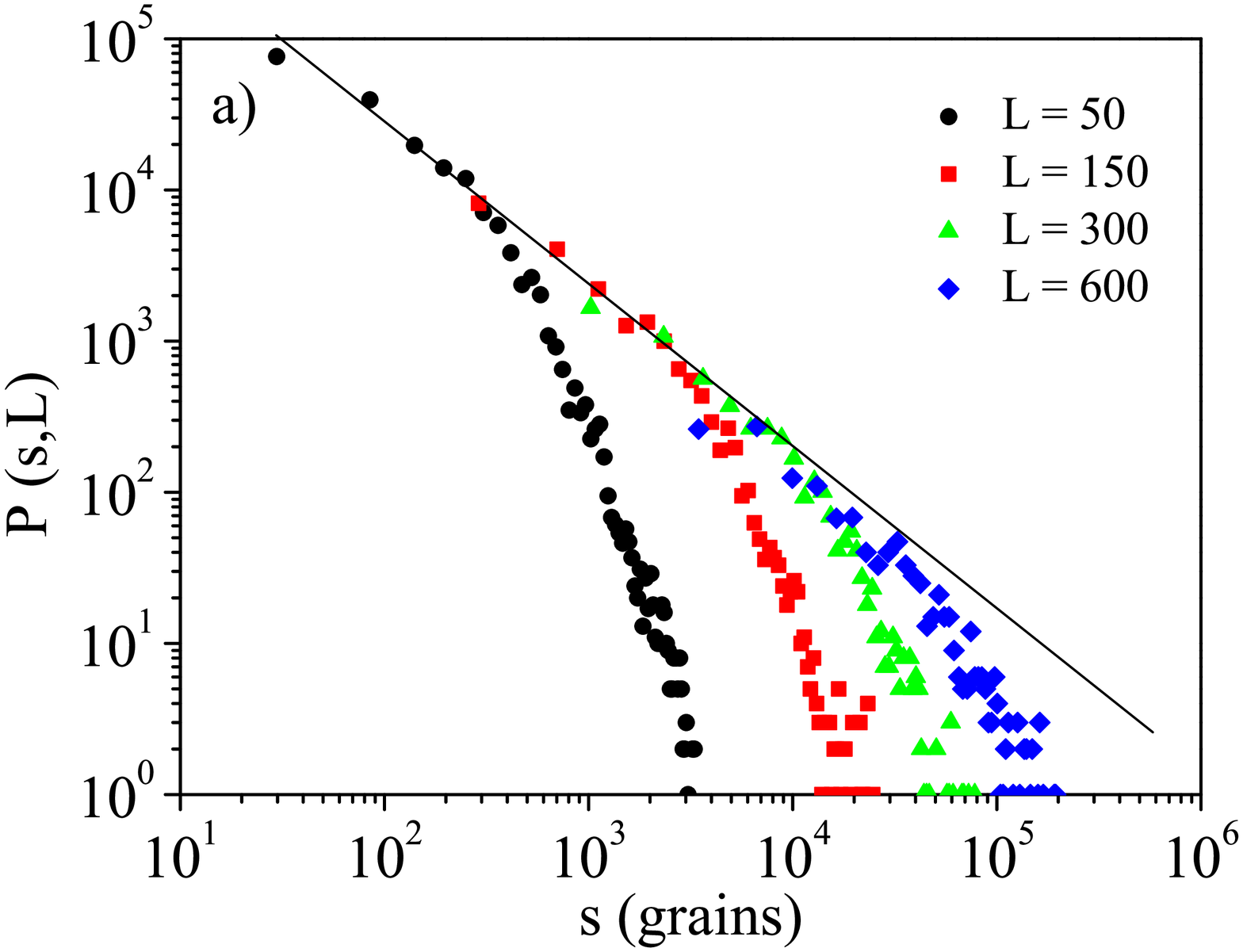}} \epsfxsize 9cm
\centerline{\epsfbox{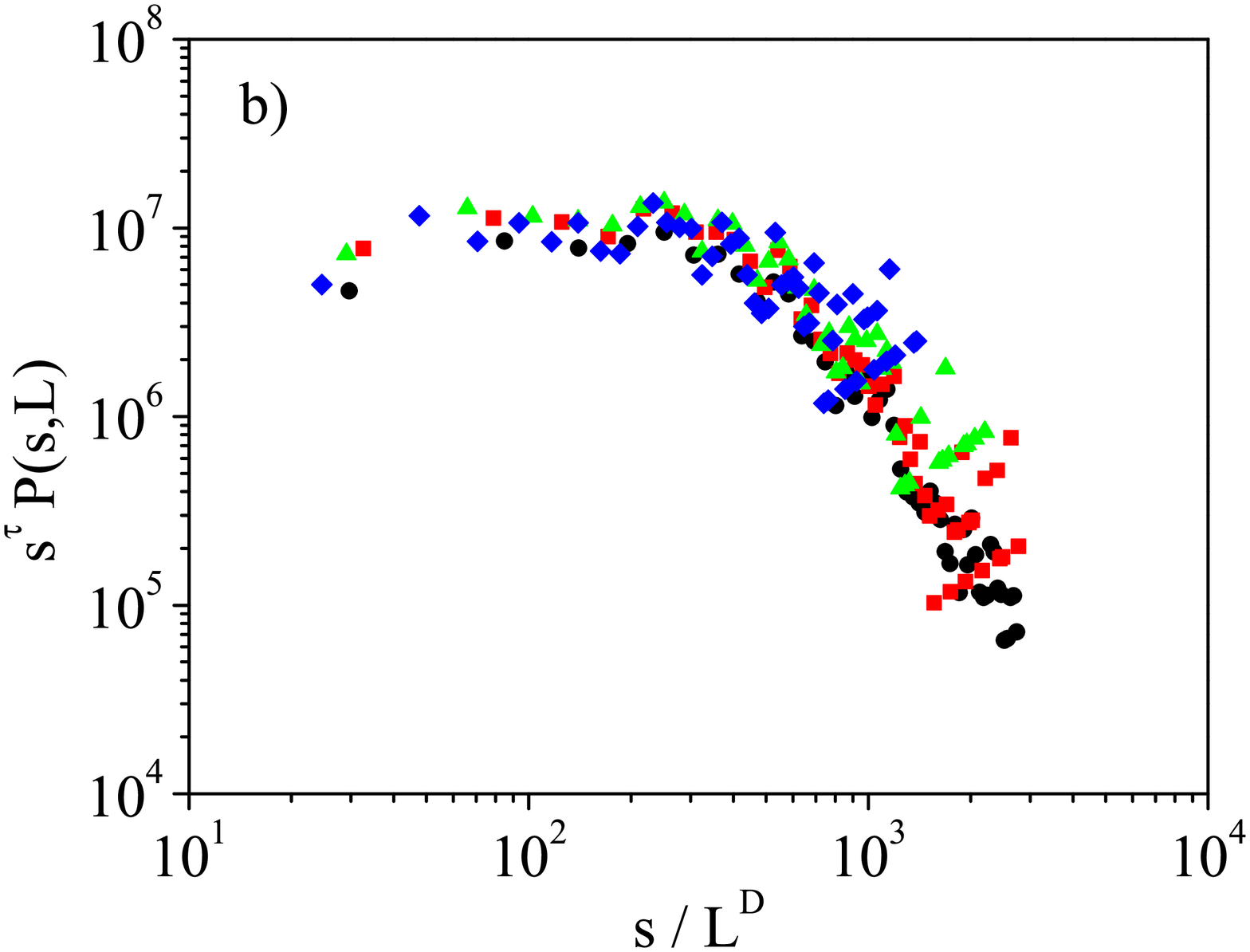}} \caption[~]{a) The unscaled size
distribution functions for the avalanches from different subsets
of the experiments corresponding to sizes of L = 50, 150, 300 and
600 mm. Given the asymptotic dependence of all them together,
there is power-law behavior over at least three decades, with an
exponent of $\tau$ = 1.20(5). b) The same data, scaled to produce
a curve collapse. The sizes of the avalanches are scaled with
$L^{-D}$ and the probabilities are scaled with $s^\tau$. The
values used to obtain the best curve collapse are $\tau$ = 1.21(2)
and $D$ = 1.99(2).} \label{hist}
\end{figure}

A histogram of these data, giving the avalanche size distribution
is shown by the diamonds in Fig.~\ref{hist}a). In that figure,
size distributions of subsets of the data corresponding to smaller
fields of view (L = 50, 150 and 300 mm) are also shown. Due to
experimental resolution, the smallest avalanches measured for each
subset depend on the size of that subset. The range of sizes from
50x50 mm$^2$ to 600x600 mm$^2$ spans more than a decade and all of
the data taken together show a power-law scaling of the size
distribution over three decades, with an exponent of $\tau$ =
1.20(5).

These data can however be scaled to fall onto a single curve, as
shown in Fig~\ref{hist}b). Here the avalanche sizes are scaled
with $L^{-D}$ and the probabilities are scaled with $s^\tau$. The
good curve collapse visible in Fig~\ref{hist}b) indicates the
presence of finite size scaling in the data, one of the hall-marks
of critical behavior in a system and thus evidence of the
appearance of SOC in the 3d rice pile.
Due to the finite size scaling found from the curve collapse, the
avalanche size distribution as a function of system size can be
written as a function of one parameter only:
\begin{equation}
P(s,L) = s^{-\tau} f(\frac{s}{L^D}),
\end{equation}
where $f(x)$ is constant up to a cutoff scale
\begin{equation}
s_{co} \propto L^D.
\label{cutoff}
\end{equation}
The exponents $\tau$ and $D$ used to obtain the curve collapse in
Fig.~\ref{hist}b) were $\tau$ = 1.21(2) and $D$ = 1.99(2). We note
here that in usual finite size scaling, separate experiments are
used instead of subsets. However, as we have tested on simulations
of a 2d version of the Oslo-model \cite{oslo2d}, finite size
scaling using subsets yields the same exponents, but is much more
easily implemented experimentally. As an additional test, the
avalanche dimension can be determined directly using a
box-counting method in 3d. This yields $D$ = 2.05(10) consistent
with the result from finite size scaling.

\subsection{Surface Roughening}

The roughness of a surface can be characterized in different ways.
Most commonly, this is done via the width of the interface as
given by its standard deviation \cite{barabasi}. For a 2d surface,
this is given by
\begin{equation}
w^2 (L) = \frac{1}{L^2} \sum_{i,j = 1}^{L} (h(i,j) - \langle h
\rangle)^2,
\end{equation}
where $\langle h \rangle$ is the average profile of the surface
height. When the roughness of the surface has reached saturation,
its value $w_{sat}$, will scale with the system size as a
power-law, $w_{sat} \propto L^\alpha$, where $\alpha$ is the
roughness exponent.

\begin{figure}
\input{epsf}
\epsfxsize 9cm 
\centerline{\epsfbox{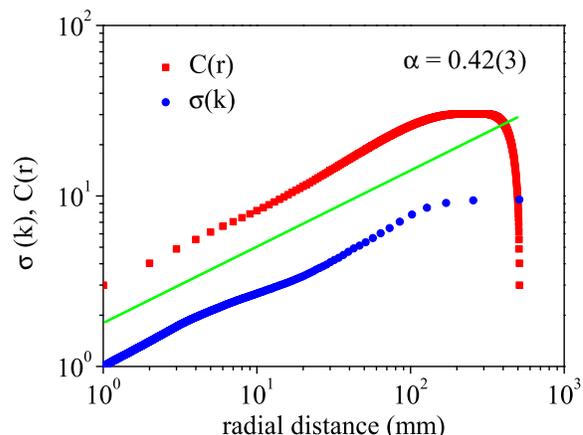}} \caption[~]{Determination of the
roughness exponent from a power-spectrum analysis as well as from
the correlation function $C(x)$ (squares). After subtracting the
average slope of the pile, its 2d power spectrum is calculated.
After taking a radial average, it is integrated over k-space in
order to give the distribution function $\sigma$, which follows a
power-law dependence $\sigma \propto k^{-\alpha}$, with a
roughness exponent $\alpha$ = 0.42(3). Similarly, the correlation
function $C(x) \propto x^\alpha$ (squares) is calculated from the
pile surface, resulting in the same exponent as that determined
from the distribution function.} \label{alpha}
\end{figure}

A similar type of analysis can be achieved via the power spectrum
of the surface. Here, one has the advantage that better statistics
can be obtained, since the whole surface can be used at once
\cite{power}. We determined the power spectrum from a radial
average of the 2d Fourier transform of the surface
\begin{equation}
S(k) = |\hat{h}(k_x,k_y)|^2,
\end{equation}
where $k = (k_x + k_y)^{1/2}$. The Fourier transforms where
performed via an FFT algorithm, which is why a data subset
corresponding to a power of two (512x512) was used. From the power
spectrum we then determine the distribution function $\sigma (k)$
\begin{equation}
\sigma^2(k) =  \int_0^k S(\kappa) \kappa d\kappa,
\end{equation}
from which it can be shown \cite{power} that $\sigma(2\pi /k) = w
(\ell)$, where $\ell = 2\pi/k$ is the length scale over which the
width is calculated. The roughness exponent $\alpha$ can therefore
be reliably obtained from a determination of $\sigma (k)$. An
alternative way of obtaining the roughness exponent is via the two
point correlation function\cite{barabasi}.
\begin{equation}
C(\vec{x},t) = \left(\langle (h(\vec{\xi},\tau) -
h(\vec{x}+\vec{\xi},t+\tau))^2
\rangle_{\vec{\xi},\tau}\right)^{1/2},
\end{equation}
the radial average of which obeys the same scaling behavior as the
distribution function \cite{barabasi}.

In Fig.~\ref{alpha}, the correlation and distribution functions
for the rice-pile surface is shown, averaged over all time steps
of all experiments (1330 images). The part of the surface studied
consists of 512x512 mm in the center of the surface laterally
(along the horizontal direction) and starting from a height of 200
mm. As can be seen from the double logarithmic plot, there is
power-law scaling with an exponent, $\alpha$ = 0.42(3).

\begin{figure}
\input{epsf}
\epsfxsize 9cm 
\centerline{\epsfbox{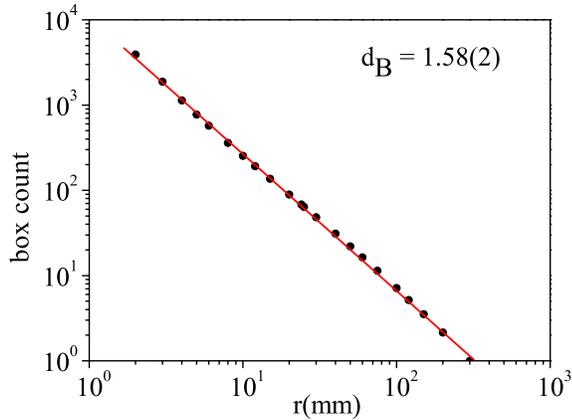}} \caption[~]{Determination of the
fractal dimension of the avalanches. Over all experiments,
avalanches with a size two standard deviations bigger than the
average were studied. The contours are derived from a thresholding
at the value of the mean height difference for each image. The
fractal dimension is subsequently determined from a simple box
counting method.} \label{frac}
\end{figure}

The scaling behavior of the interface width can, however, also be
determined from the cut-off size of the avalanches in a system
with size $L$. The size of such an avalanche will be proportional
to the maximal height difference, given by the saturation width
$w_{sat}$, times the maximal area of the avalanche. As can be seen
from Fig.~\ref{diff}, the avalanche-shape is a fractal, which
means that $s_{co} \propto w_{sat}L^{d_B}$, where $d_B$ is the
fractal dimension of the cluster of displaced grains. Equating
this expression with Eq.~\ref{cutoff}, one obtains $w_{sat}
\propto L^{D-d_B}$, which implies the scaling relation
\begin{equation}
\alpha = D - d_B. \label{alpha2}
\end{equation}
Such a scaling relation between the roughness exponent and the
exponents characterizing the avalanches was also derived by
Paczuski {\em et al.} \cite{pacz} from different arguments.

In order to determine $d_B$ in an accurate manner, we studied
avalanches more than two standard deviations bigger than the
average size. This corresponds to a subset of $\sim$ 100 images
from all experiments. After thresholding the height difference
fields $\Delta h(x,y)$ at their mean value, we applied a simple
box counting method \cite{mandelbrot} to the resulting clusters.
This is shown in Fig.~\ref{frac}, where the number active pixels
in a box of given size is shown as a function of the box size. The
result is a power law with an exponent of -1.58(2), indicating a
fractal dimension of $d_B$ = 1.58(2). Inserting this value into
scaling relation (\ref{alpha2}), the roughness exponent is found
to be $\alpha$ = 0.41(3), in excellent agreement with that
determined from the roughness analysis.

\begin{figure}
\input{epsf}
\epsfxsize 9cm 
\centerline{\epsfbox{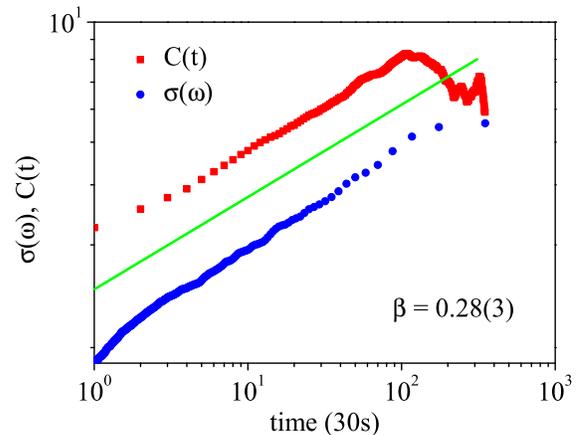}} \caption[~]{Determination of the
growth exponent from a power-spectrum analysis. Here, the
distribution function $\sigma$ has been determined from the time
dependence of the height of each pixel. The data show a power-law
dependence $\sigma \propto \omega^{-\beta}$, with a growth
exponent $\beta$ = 0.28(3). The same result is obtained via the
correlation function, indicated by the open symbols. From the
growth and roughness exponents, the dynamic exponent $z =
\alpha/\beta$ = 1.5(1) can be determined.} \label{beta}
\end{figure}

The dynamics of the roughening process is likewise analyzed via
the distribution function. In contrast to the roughness analysis
above, we now determine the distribution function for the time
dependence $h_{i,j} (t) - \langle h_{i,j} \rangle_t$ for each
pixel, where $\langle \cdot \rangle_t$ denotes the average over
the duration of the experiment. Again, the distribution function,
given by
\begin{equation}
\sigma^2(\omega) = \int |\hat{h}(\omega)|^2 d\omega,
\end{equation}
is equal to the momentary width, $\sigma (2\pi/\omega) = w(t)$.
Thus the growth exponent $\beta$, describing the scaling of the
width with time, can be determined from the distribution function.
Again, the correlation function shows the same scaling behavior in
time, allowing a separate determination of $\beta$. Both results
are shown in Fig.~\ref{beta}, where it can be seen that there is
good power-law scaling of the distribution function, as well as
the correlation function, over two decades with an exponent of
$\beta$ = 0.28(3). Another way of describing the dynamics of an
interface is via the dynamic exponent $z$, which describes the
scaling of the saturation time with the system size
\begin{equation}
t_\times \propto L^z. \label{tx}
\end{equation}
It can be easily obtained \cite{barabasi} from the roughness and
growth exponents via
\begin{equation}
z = \frac{\alpha}{\beta}.
\end{equation}
From the values determined above we thus obtain $z$ = 1.5(1).

The dynamic exponent, $z$, can also be derived from a scaling
relation using the critical exponents of the avalanche dynamics.
The saturation time, $t_\times$, will be roughly given by the time
it takes for an avalanche of the cut-off size, $s_{co}$, to
appear. Since material is added to the system at a constant rate,
the number of grains added until a cut-off avalanche occurs is
proportional to the cross-over time $t_\times$. On the other hand,
the material added not only result in an avalanche of size
$s_{co}$, but will also be lost by smaller avalanches, such that
the total material necessary to obtain an avalanche of size
$s_{co}$ can be estimated from:
\begin{equation}
\int_0^{s_{co}} s P(s) ds \propto s_{co}^{2-\tau}.
\end{equation}
From Eq.~\ref{cutoff}, we obtain for the cross-over time $t_\times
\propto L^{D(2 - \tau)}$. With Eq.~\ref{tx}, this immediately
leads to the scaling relation
\begin{equation}
z = D(2 - \tau),
\end{equation}
which was also derived by Paczuski {\em et al.} \cite{pacz} from
more general arguments. Using the values for $D$ and $\tau$
determined above from the curve collapse of the avalanche size
distributions, we obtain $z$ = 1.56(8), which is again in very
good agreement with the result of the power spectrum analysis, $z$
= 1.5(1). Note also that the exponents $\alpha$ and $z$ fulfill
the KPZ scaling relation\cite{KPZ} $\alpha + z$ = 2, which is
valid independent of the dimension of the system and depends only
on the fact that the growth is driven by height gradients.
Furthermore, the roughness exponent $\alpha$ = 0.42(3) is in good
agreement with a numerical determination of the behavior of the 2d
KPZ equation \cite{2dKPZ}. The connection between roughening and
SOC will be discussed in more detail below.

\section{Conclusions}

In conclusion, we have presented results on both the avalanche and
the roughening behavior of an experimental SOC system. In
addition, we presented simple arguments for universal scaling
relations, derived by Paczuski {\em et al.} \cite{pacz} on general
grounds, connecting the avalanche and roughening behavior. We
obtain {\em quantitative} agreement of the experimental exponents
characterizing the roughness and those describing the avalanche
statistics. This means that the roughening of the surface of the
pile is governed by its underlying avalanche dynamics, which was
already conjectured from models of interface depinning. Earlier
experimental studies \cite{Frette,rice} have considered the
avalanche dynamics and the roughening behavior of the 1d version
of the present system, where finite size scaling was also found
\cite{Frette}. The above scaling relations, showing quantitatively
the connection between the two phenomena were not previously
tested.

The fact that KPZ dynamics is observed in a SOC system, as
indicated by the fact that the roughness and dynamic exponents
obtained here fulfill the KPZ scaling relation, is surprising.
This is mostly because in SOC systems, the critical state is self
organized, whereas in KPZ systems, the roughening is put into the
dynamics by obeying the proper symmetries. However, it has been
argued by Alava {\em et al.} \cite{alava}, that there is a mapping
of SOC models, which naturally lead to the necessary symmetries.
In this context, it should also be noted that experiments on KPZ
systems such as burning paper have also observed 'avalanches' in
the advance of fronts \cite{jussi2}.

\section{Acknowledgements}

This work was supported by FOM (Stichting voor Fundamenteel
Onderzoek der Materie), which is financially supported by NWO
(Nederlandse Organisatie voor Wetenschappelijk Onderzoek).

\bibliographystyle{prsty}

\end{multicols}
\end{document}